# Anomaly in the dielectric response at the charge orbital ordering transition of $Pr_{0.67}Ca_{0.33}MnO_3$

Silvana Mercone, Alexandre Wahl, Alain Pautrat, Michaël Pollet and Charles Simon
Laboratoire CRISMAT, CNRS UMR 6508, ENSICAEN-CNRS
6 Bd. du Maréchal Juin, 14050 Caen Cedex, France

The complex impedance of a $Pr_{0.67}Ca_{0.33}MnO_3$ crystal has been measured. The frequency dependence is studied for a wide range of temperatures (50K-403K) and is found to be characteristic of relaxation process with a single Debye time relaxation constant, which is interpreted as a dielectric constant of the material. A strong peak is observed in this dielectric constant (up to $10^6$) at the charge ordering transition suggesting an interpretation in terms of ordering of electric dipoles at $T_{CO}$ or in term of phase separation. Comparison with $Pr_{0.63}Ca_{0.37}MnO_3$ - in which the phase separation is much smaller and the peak in the dielectric constant is absent - suggests an interpretation in term of phase separation between insulating and metallic states.

## I.INTRODUCTION

Mixed-valent manganites have been the focus of intense scientific activity over the past several years in view of the variety of physical phenomena they display as well as their potential for utilization in magnetic sensing and spin-polarized transport applications.[1]

Of particular interest is the study of the behavior of the ordered state in these compounds. The competition between the charge-ordered (CO) insulating state, in which the electric charges are localized, and the charge-delocalized (CD) state, which presents a metallic-like conductivity, is of a great attention.[2,3,4] The physics of this electronic phase transition are also linked to the magnetism of the system. The CO state is stabilized by an antiferromagnetic (AFM) ground state while that of the CD state is ferromagnetic (FM).[5-7] Moreover, many experiments have shown that the CO state is unstable under a variety of external perturbations including magnetic field,[8-10] temperature, external and chemical pressure[11] (i.e. degree of cation/anion doping), x-ray[12] and electron[13] irradiation.

Among the various mixed-valence manganites studied so far, the $Pr_{1-x}Ca_xMnO_3$ (PCMO) system is of great interest for the study of the competition between the CO and CD states previously mentioned. Indeed, the similar ionic radii of the A-site cations limits the distortion of the perovskite lattice upon doping.

In the last few years, experiments have focused on the effects of the application of an electric field on doped manganites. Results for the system $Pr_{1-x}Ca_xMnO_3$ with doping x=0.3, 0.33, 0.4 are similar for bulk crystals,[14] single crystals,[15,16] ceramic samples[17] and thin films,[18,19]. A drop in the resistance above a threshold value of the applied DC-electric field (or DC-current) is generally observed. This electric-field-driven (or current-driven) insulator to metal transition is associated to a non-linear conductivity. The non-linearity of the I-V characteristics observed for $T < T_{CO}$ is explained by Guha *et al.*[15,16] considering a percolation process due to the melting of the CO-Insulating (COI) phase into the CD-Metallic one. This interpretation is based on the opening of metallic-filament paths in a COI-matrix. Stankiewicz *et al.*[17] have rejected this one-dimensional explanation in manganites, as it is incompatible with the magnetic data they obtained under the application of a current. A filamentary picture of conductivity should result in a change in magnetization of only 0.05%. However, they observed a much larger variation. Thus they have invoked a growing of FM-volumic-clusters along the direction of injected-current that would be at the origin of such a percolation-based scenario. A more recent paper[20] has also shown the presence of the same non-linearity in a non-charge-ordered composition (x=0.2) in which the percolation model was not applicable adding complexity to this features.

To add to this scenario of differing views, recent spectroscopy experiments[21,22] have suggested the development of a charge-density-wave (CDW). This is supported by the quasi-one-dimensional electronic structure proposed by Asaka *et al.*[23] in the light of their low-temperature TEM measurements. A.Wahl *et al.*[24] have also proposed the CDW theory in order to explain the I-V behavior of $Pr_{0.63}Ca_{0.37}MnO_3$. This phenomenological model was originally limited to the compound $NbSe_3$, [25-30] but was later generalized to a number of transition-metal compounds[31] in order to explain the non-linear effects observed in their ac and dc transport properties. To complete the panorama of studies on this competition between the COI-AFM and CD-FM states, Yamada *et al.*[32] have studied the dielectric spectra for different doping and as a function of the frequency for the low temperature range (40K<T<130K). They proposed a small-polaron-hopping model to describe both dc and ac transport. The authors have found the maximum value of the activating energy for the composition x≈0.33. On the other hand the work of Rivadulla *et al.*[33,34] focused on the high temperature region where they found a large capacitive response of the material at the occurrence of the CO regime. They suggest, in the light of their magnetization measurements, the existence of a strong FM-AFM competition for $T<T_{N2}$=160 K. As a consequence of AFM order in the CO state, hopping between the Mn sites is minimized so that the formation of the electric dipoles is permitted.

The aim of this report is to elucidate the nature of the electric-field induced conducting states of PCMO. We report complex impedance measurements of $Pr_{0.67}Ca_{0.33}MnO_3$ crystal over four orders of magnitude of frequency and over a large range of temperatures (50K-400 K). The real part of the ac resistance (ReZ) shows a broad decrease at a crossover frequency which is highly dependent on temperature and at which the imaginary part of the impedance (ImZ) exhibits a peak. We have succeeded in reproducing this typical relaxation behavior using a Debye model. The relaxation time is strongly dependent on temperature, and the corresponding dielectric constant displays an anomaly at $T_{CO}$.

## II. EXPERIMENTAL

Using the floating-zone method with feeding rods of nominal composition $Pr_{0.7}Ca_{0.3}MnO_3$, a several cm-long single crystal was grown in an image furnace. The sample for measurement was cut from the middle part of this crystal**.** The electron diffraction (ED) investigation showed the existence of twinning domains. Therefore, all physical measurements performed on such samples are averaged over the six oriented domains coexisting in the Pnma phase. Electron Dispersive Spectroscopy analysis found the cationic composition to be x=0.33. Magnetization and dc-transport measurements were carried out in a SQUID magnetometer and in a PPMS-Quantum Design cryostat, respectively. Under zero magnetic field, the transport measurements show a classical insulating behavior with a resistivity up to $10^9$ Ωcm at 5 K. A change of slope can be observed at the onset of the charge ordering state. The characteristic temperatures are: $T_{CO}$ about 225 K, $T_C$ about 100 K and $T_N$ about 115 K.[35]

The low temperature complex impedance (T<250K) was measured by means of a lock-in amplifier in the frequency range (1Hz-$10^5$Hz). Two different circuits were used depending on the investigated range of temperature (below and above 100K). With this experimental set up no relevant frequency dependence of the complex impedance could be put in evidence for T>130K. This point is developed in the following.

In the high temperature range (210K<T<403K) the dielectric constant was measured using a LCR Bridge (Flucke 6306). Due to the limits of the setup, the lowest possible temperature for these measurements was 210 K. The sample was painted on both faces with In-Ga paste, and a 1 V AC-voltage was applied in the (50-1*$10^6$) Hz range. The contributions of the measured impedance, the air permittivity, the contacts and the cables (resistive and inductive effects) were carefully removed in order to obtain the actual permittivity of the sample.

## III.RESULTS

In Figure 1, the modulus of $Z = \sqrt{ReZ^2 + \mathrm{Im}\,Z^2}$ as a function of temperature is shown for frequencies ranging from 10 to $10^5$ Hz. As previously mentioned, the impedance of the sample varies strongly (several orders of magnitude) over the investigated range of temperatures; thus two circuits, depending on the temperature range, were used to carry out our measurements. The limit of validity of each circuit is denoted by the dashed line. The inset on the left side of Figure 1 shows the circuit for the range 80K-120K. A constant voltage is applied and saturates when the resistance of the sample falls below R (R represents the saturation value of the circuit, R=100MΩ). The inset on the right side of Figure 1 shows the circuit used in the range 120K-250K**.** In this case a constant value of the current can be injected only if the resistance of the sample remains much lower than R. Solving in both cases the equivalent circuit, it is possible to extract the complex impedance of the sample. This kind of behavior has already been observed by J. Sichelschmidt *et al.*[36] on a related material. However, no interpretation was given to account for the data.

Figure 2(a) shows the modulus $Z = \sqrt{ReZ^2 + \operatorname{Im} Z^2}$ of the complex impedance $\overline{Z}$ within the range 80K-120K as a function of frequency. A decrease of many orders of magnitude occurs when the frequency is increased. Both the crossover frequency $\nu_C$ (defined as the frequency where $Z = \frac{1}{\sqrt{2}} R_{DC}$) and the low-impedance value ($Z(\omega \rightarrow 0) = R_{DC}$) are clearly temperature-dependent. As it can be seen in figure 2(a), the frequency range explored is not sufficientlly large to detecte the crossover frequency for highest temperature. This is the reason why we can not show data between 140K and 210K. Above 210K up to 403K, we used the LCR Bridge.

Figure 2(b) displays the frequency dependence of Z investigated at high temperature (210K-403K). Several features are noteworthy. The most important is the appearance of an inductive term (Z proportional to ω) nearly independent from temperature that we will interpret in term of induction in the wiring.

## IV.ANALYSIS AND DISCUSSION

The above results are characteristic of a relaxational process[32,33,34]. The real and imaginary parts of $\overline{Z}$ present a crossover frequency at which the former starts to decrease and the latter exhibits a concurrent peak (Figures 3 and 4). This type of behavior has been quantitatively explained in CDW systems (such as $NbSe_3$ or the “blue bronze”) as a consequence of the excitation of a collective mode (CM) [25-31] , but can be also observed in presence of a large dielectric constant which is often fitted by a resistance and a capacitor in parallel in the equivalent circuit. [24-34]

In all these measurements, the ac current intensity was chosen small enough to prevent any non linear or heating effects often observed in DC measurements. In light of these considerations it is possible to model the observed ac-response with the equivalent electronic circuit shown in Figure 5(a). The overall sample $\overline{Z}$ can be represented using a circuit where the resistance $R_N$ is in parallel with an additional element that defines the capacitive response of the sample. In the limit $\beta = \frac{R_{CM}}{R_N} \to 0$, we obtain the new linear-equivalent-circuit shown in Figure 5(b). The real and imaginary parts of the sample impedance are:

$$Z_{real}(\omega) = \frac{R_N}{1+\tau_0^2\omega^2} \qquad (1)$$

$$Z_{imaginary}(\omega) = -[Z_{real}(\omega) * \tau_0\omega] \qquad (2)$$

where $\tau_0 = R_N C$

From the above expressions, one can note that the response obtained is that of a Debye relaxation process where the equivalent linear circuit is given by an RC one. At low frequency, the limit is a constant value (R ) and in the high frequency limit, it is 1/Cω, as it is clearly observed on figure 2 (a).

Using Equation (1), we fit the real part of $\overline{Z}$ (ω) at various temperatures and then plug the obtained fit parameters ($R_N$ and C) into Equation (2) to reproduce the imaginary part of $\overline{Z}$ (ω). As seen in Figures 4 and 5 there is a good agreement between the model and the experimental data and this is observed within the whole temperature range. The open circles in Figure 6 represent the relaxation time $\tau_0$=$R_N$C as calculated by using the obtained fitting parameters, $R_N$ and C, within the Debye model description. Contrary to Yamada *et al.*[32] no empirical statistical distribution is needed to adjust the model to the experimental data. At high temperature (T>250K) an inductive term is necessary to account for the data (see figure 2). This inductive contribution, arising from contact leads (usually observed at higher frequency[37] (>10 MHz)), is constant in the whole range of temperature (L≈5.57 µHenry). For T>250K the capacitive term decreases so the inductive L starts to be relevant to account for the experimental data. When the temperature decreases below 250K, the value of C increases so the inductive term starts to be negligible and the analysis by RLC circuit gives the same capacitance values as the one done by using the classical RC circuit. This allows to say that this change in the circuit used for the analysis cannot cause the observed anomaly at $T_{CO}$.

In order to obtain more information about the temperature dependence of τ, let us return to Equation (1) and use the $\overline{Z}$ measured at different frequencies as a function of the temperature. Rearranging the Equation (1) and applying it for frequency $\omega_{HF} >> \omega_{LF}$ results in:

$$\tau_0(T) = \frac{1}{\omega_{HF}}\sqrt{\frac{R_N(T)}{Z_{real,\omega_{HF}}(T)} - 1} \qquad (3)$$

Through Equation (3) it is possible to reconstruct the low temperature dependence of the relaxation time (open circles in Figure 6). The high temperature dependence (T>160K) of the relaxation time curve cannot be obtained this way since Equation (3) is only valid for $\frac{Z_{real}(\omega_{LF})}{Z_{real}(\omega_{HF})} \neq 1$. As seen in Figure 6, the resulting values (closed squares) are in good agreement with the parameters of the fit in the same range of temperature (open circles within 50K and 120K). From the semi-log scale in Figure 6, it is apparent that the observed process does not exhibit the same temperature-activated law through the whole range of T. In the temperature regime 210 K-400 K, the Arrhenius law is obeyed with the activating energy $E_a \approx 31.14$ meV, which is of the same order of magnitude as that found by Rivadulla *et al.* for the same temperatures.[33,34] In the low temperature region (80 K-170 K), the relaxation time does not follow an activated law and a peak at the charge ordering temperature is observed. At low temperature (T<100K) the relaxation time starts to saturate. The behavior shown in figure 6 dramatically illustrates the strong temperature dependence of the relaxation time constant which is not a trivial feature.

In order to shine light on the role of the magnetic and/or charge ordered state on the capacitive intrinsic response of PCMO, we have transformed the complex impedance into a dielectric constant. The analysis developed above gives the values of the capacitance of the PCMO as a function of the temperature. The permittivity is calculated through the relation $\varepsilon = \frac{Ce}{s\varepsilon_0}$, e being the thickness of the sample, $\varepsilon_0$ the vacuum dielectric constant (8.85 pF/m) and s the electrode surface. The obtained dielectric constant is shown as a function of the temperature in Figure 7. The first important feature of these curves is that there is a peak in the $\varepsilon(T)$ at $T_{CO}$=225 K. The second one is the absence of any large anomaly in $\varepsilon(T)$ at the magnetic transitions ($T_N$=115K, $T_C$=100K). It can be also noticed that the dielectric constant is not zero in the low temperature phase, whereas it is vanishingly small in the high temperature limit.

All these features strongly suggest that we are in presence of a electric dipolar transition at $T_{CO}$ with a diverging permittivity at the phase transition. Let us now come back on the exact nature of the superstructure of such compounds. Up to now the CO state was considered only as a site-centered ordering (or metal-centered or Mn-centered) (Figure 8 (a)). This interpretation is the classical one in which one consider the alternation of electron occupancy on Mn ions and which, for the half-doped material, presents the well-known CE-type magnetic structure[38,39]. However, in recent papers[40,41], this picture was questioned, since another type of structure was found to be more consistent with the experimental data. The CO state is considered in this case as an oxygen-centered ordering (or bond ordering as it concerns the bonding interaction Mn-O-Mn). In this model, the Mn oxidation state is constant (+3,5) and the supernumerary electron associated to the Double Exchange (DE) Zener mechanism, ferromagnetically couples the Mn ions in the pair originating in the so-called Zener Polaron (Figure 8 (b)).

Komskii *et al.*[42] suggested that this Zener polaron state may be favored in certain part of the phase diagram of PrCaMnO systems. In this case, a net polarization would appear favoring a ferroelectric state (Figure 8 (c)). Within such a scenario, an anomaly on the dielectric constant at the charge ordering temperature is expected. On the contrary, recent analysis of the diffraction data[44], in agreement to what was previously published[40], state that no ferroelectric dipole exists in this structure, letting the possibility of electric antiferroelectric transition with a divergence of the dielectric constant[45]. This should be more impressive in the $Pr_{1/2}Ca_{1/2}MnO_3$ (where no single crystal exists) or in $Pr_{0.63}Ca_{0.37}MnO_3$ in which the distortion due to the Zener polarons is larger. Our measurements on $Pr_{0.63}Ca_{0.37}MnO_3$ failed to exhibit any measurable dielectric constant at low temperature (the drop of the resistivity does not occur up to the largest studied frequency). This rather suggests that the origin of the dielectric constant is due to the mixing of metallic and insulating phases which was evidenced in $Pr_{0.66}Ca_{0.33}MnO_3$ in zero magnetic field by different techniques below 100K[35]. The present results also suggest that the phase separation also exists above the magnetic transitions temperatures 100K up to $T_{CO}$, since another important feature in Figure 7 is the absence of any anomaly on the permittivity behavior at these magnetic transition temperatures.

The colossal capacitive response is due in this model to the presence of a very large interface between the two phases as is was observed by small angle neutron scattering[35]. If one tries to estimate the capacitance from this specific surface ($C=\varepsilon_0 S/t$), one can calculate the capacitance per unit of volume V: $C/V = \varepsilon_0 (S/V)/t = = \varepsilon_0/t^2$ . The thickness t of the insulating phase extracted from neutron data is about 1nm, so C/V can be really colossal, up to $10^9$ F/m$^3$. Moreover, the quantity of interface strongly increases near $T_{CO}$, as it was recently observed by neutron small angle scattering[46] and this can explain the peak in the capacitance. However, there are many difficulties to apply such a simple formula in such a nanoscopic phase separation proposed in reference 35.

## V. CONCLUSION

In conclusion the complex impedance has been measured in a $Pr_{0.67}Ca_{0.33}MnO_3$ crystal. Its frequency dependence, observed under zero magnetic field for a wide range of temperatures, is characteristic of a relaxation process with a single Debye time relaxation constant. On the one hand, this type of behavior can be interpreted in the framework of a CDW condensate depinning as the equations generally used to explain the transport in a CDW system are very general and can be used for any kind of system which present a strong capacitive response. On the other hand this kind of behavior is exactly the same observed in the polarized systems where the condensate is due to electric dipoles or to the presence of a very large interface between insulating and metallic parts. The anomaly observed in the dielectric constant at $T_{CO}$ and the absence of any similar peak in non phase separated systems, is in agreement with an interpretation in term of phase separation. This opens a new research area in the domain of colossal dielectric constant materials[47].

## VI. ACKNOWLEDGMENTS

We acknowledge L. Hervé and M. Strebel Morin for sample preparation and D. Khomskii, M.B. Weissman and V. Caignaert for useful discussions we had the opportunity to have with them. S. Mercone acknowledges support from European Community. We thank our referee for pointing out the influence of the inductive term on the analysis of the temperature dependence of the relaxation time, which makes the present version of the paper much clear. We acknowledge the "région basse Normandie" for financial support of the experimental setup.

# REFERENCES

[1]For a review, see for example *"Colossal Magnetoresistive Oxides"*, edited by Y.Tokura (Gordon and Breach Science, New York, 1998), *"Metal-insulator transitions"*, M.Imada, A.Fujimori and Y.Tokura (Reviews of Modern Physics Oct. 1998), *"The Physics of manganites: Structure and transport"*, M.B.Salamon and M.Jaime (Reviews of Modern Physics, July 1998) and *"Colossal Magnetoresistant materials: the key role of phase separation"*, E.Dagotto, T.Hotta and A. Moreo (Physics-Reports, April 2001).

[2]P.Schiffer, A.P. Ramirez, W. Bao, S.-W. Cheong, Phys. Rev. Lett. **75,** 3336 (1995).

[3] A. Urushibara, Y. Moritomo, T. Arima, A. Asamitsu, G. Kido, Y. Tokura, Phys. Rev. B **51,** 14103 (1995).

[4] G. Varelogiannis, Phys. Rev. Lett. **85,** 4172 (2000).

[5] T. Hotta and E. Dagotto, Phys. Rev. B **61,** R11879 (2000).

[6] I.V. Solovyev, Phys. Rev. B **63,** 174406 (2001).

[7] S.Shimomura, T. Tonegawa, K. Tajima, N. Wakabayashi, N. Ikeda, T. Shobu, Y. Noda, Y. Tomioka and Y. Tokura, Phys. Rev. B **62,** 3875 (2000).

[8] H. Yoshizawa, H. Kawano, Y. Tomioka, Y. Tokura, Phys. Rev. B **52,** R13145 (1995).

[9] A. Anane, J.-P. Renard, L. Reversat, C. Dupas, and P. Veillet, M. Viret, L. Pinsard and A. Revcolevschi , Phys. Rev. B **59,** 77 (1999).

[10] Y. Tomioka *et al.*, Phys. Rev. B **53,** R1689 (1996).

[11] Y. Moritomo, H. Kuwahara, Y. Tomioka and Y. Tokura, Phys. Rev. B **55,** 7549 (1997).

[12] V. Kiryukhin *et al*, Nature (London) **386**, 813 (1997).

[13] M. Hervieu, A. Barnabé, C. Martin, A. Maignan and B. Raveau, Phys. Rev. B **60,** R726 (1999).

[14] A. Asamitsu, Y. Tomioka, H. Kuwahara and Y. Tokura, Nature (London) **388**, 50 (1997).

[15] A.Guha, A.K. Raychaudhuri, A.R. Raju and C.N.R.Rao, Phys.Rev.B **62,** 5320 (2000).

[16] A.Guha, N.Khare, A.Raychaudhuri and C.N.R.Rao, Phys.Rev.B **62,** R11941 (2000).

[17] J. Stankiewicz, J. Sesé, J. Garcia, J. Blasco and C. Rillo, Phys.Rev.B **61,** 11236 (2000).

[18] S. Srivastava, N. K. Pandey, P. Padhan and R.C. Budhani, Phys.Rev.B **62,** 13868 (2000).

[19] C.N.R. Rao, A.R. Raju, V.Ponnambalam, S.Parashar and N.Kumar, Phys.Rev.B **61**, 594 (2000).

[20] S. Mercone, A. Wahl, Ch. Simon and C. Martin, Phys.Rev.B **65**, 214428 (2002).

[21] N. Kida and M. Tonouchi, Phys.Rev.B **66**, 024401 (2002).

[22] P. Calvani, G. De Marzi, P. Dore, S. Lupi, P. Maselli, F. D'Amore, S. Gagliardi and S-W. Cheong, Phys. Rev. Lett **81**, 4504 (1998).

[23] T. Asaka, S. Yamada, S. Tsutsumi, C. Tsuruta, K. Kimoto, T. Arima and Y. Matsui, Phys. Rev. Lett. **88**, 097201 (2002).

[24] A. Wahl, S. Mercone, A. Pautrat, M. Pollet, Ch. Simon and D. Sedmidubsky, cond.-mat/0306161

[25] P.Monceau, J. Richard and M. Renard, Phys. Rev. B. **25**, 931 (1982).

[26] J. Richard, P.Monceau and M. Renard, Phys. Rev. B. **25**, 948 (1982).

[27] G. Grüner, A. Zawadowski and P.M. Chaikin, Phys. Rev. Lett. **46**, 511 (1981).

[28] R.M. Fleming and C.C. Grimes, Phys. Rev. Lett. **42**, 1423 (1979).

[29] P.Monceau, N.P. Ong, A.M. Portis, A. Meerschaut and J. Rouxel, Phys. Rev. Lett. **37**, 602 (1976).

[30] N.P. Ong and P.Monceau, Phys. Rev. Lett. **16**, 3443 (1977).

[31] R.J. Cava, R.M. Fleming, P. Littlewood, E.A. Rietman, L.F. Schneemeyer and R.G. Dunn, Phys. Rev. B **30**, 3228 (1984).

[32]S. Yamada, T.-H. Arima and K. Takida, J.Phys. Soc. Jpn. Vol 68, 3701 (1999).

[33] F.Rivadulla, M.A. Lopez-Quintela, L.E. Hueso, C. Jardon, A. Fondado, J. Rivas, M.T. Causa and R.D. Sanchez, Solid State Communications 110 (1999) 179-183.

[34] C. Jardon, F.Rivadulla, L.E. Hueso, A. Fondado, M.A. Lopez-Quintela, J. Rivas, R. Zysler, M.T. Causa and R.D. Sanchez, J. Mag.Mag.Mat. 196-197 (1999) 475-476.

[35]Ch. Simon, S.Mercone, N. Guiblin, C. Martin, A. Brulet and G. André, Phys. Rev. Lett. 89, 207202 (2002)

[36] J. Sichelschmidt, M. Paraskevopoulos, M. Brando, R. Wehn, D. Ivannikov, F. Mayr, K. Pucher, J. Hemberger, A. Pimenov, H.-A. Krug von Nidda, P. Lunkenheimer, V. Yu. Ivanov, A. A. Mukhin, A. M. Balbashov and A. Loidl, Eur. Phys. J. B **20**, 7-17 (2001).

[37]Z. Jirak, S. Krupicka, Z. Simsa, M. Dlouha and S. Vratislav, J. Mag.Mag.Mat. 53, 153 (1985)

[38]M.E.Lines, Phys.Rev.B **19**, 1189 (1979)

[39]P.G. Radaelli, D.E. Cox, M. Marezio and S.W. Cheong, Phys.Rev. B 55, 3015 (1997).

[40] A. Daoud-Aladine, J. Rodriguez-Carvajal, L. Pinsard-Gaudart, M. T. Fernández-Diaz, and A. Revcolevschi, Phys. Rev. Lett. 89, 097205 (2002).

[41]A. Daoud-Aladine, J. Rodriguez-Carvajal, L. Pinsard-Gaudart, M. T. Fernández-Diaz, and A. Revcolevschi, cond.-mat/0111267 (16 November 2001)

[42] D. Khomskii, et al. cond mat 0306651.

[43] T. Tybell, P. Paruch, T. Giamarchi, J.M. Triscone, Phys. Rev. Lett. 89, 097601 (2002).

[44] N. Guiblin and D. Grebille, private communication.

[45] An antiferroelectric phase can be obtained with two (or more) compensated dipoles in each unit cell, providing no net polarization in the ordered phase, but a possible divergence of the susceptibility, similar to what is observed in antiferromagnetic phases.

[46] D. Saurel, Ch. Simon, C. Martin, A. Brulet, private communication.

[47] C.C. Homes, T. Vogt, and S. M. Shapiro, S. Wakimoto, M. A. Subramanian, A. P. Ramirez, Phys. Rev. B 67, 092106 (2003).

## FIGURE CAPTIONS

**Figure 1**: The modulus $Z = \sqrt{ReZ^2 + \mathrm{Im}\,Z^2}$ of the complex impedance of the sample as function of temperature at different frequencies. Left inset: the circuit used at low temperature, which applies at constant voltage and saturates when the resistance of the sample and R are of the same order of magnitude. Right inset: the circuit used at high temperature, which applies at constant current if the resistance of the sample is lower than R.

**Figure 2**: The modulus Z of the complex impedance as a function of the frequency at (a) low T (b) and high T. The threshold frequency and the impedance in the DC limit ($Z(\omega\rightarrow 0)$) clearly change as a function of temperature. The behavior of Z changes drastically between 253 K and 273 K.

**Figure 3**: Real part of the complex impedance $\overline{Z}$ as a function of the pulse $\omega$ at T=80 K. Open circles are the experimental data and line corresponds to the fit from Equation (1).The values of the parameters found are: $\tau_0=(1.6*10^{-4})$ sec$^{-1}$ and $R_N\approx(3.6*10^5)\ \Omega$.

**Figure 4**: Imaginary part of the complex impedance $\overline{Z}$ as a function of the pulse $\omega$ at T=80 K. Open circles are the experimental data and the line corresponds to the simulation by Equation (2) using the parameters from the fit of the $Z_{real}(\omega)$ (figure 3).

**Figure 5**: (a) Equivalent circuits used to simulate the electrical response of the sample : $R_N$ = resistance of the normal electrons, $R_{CM}$ and $C_{CM}$=, resistance and capacity respectiveley.

(b) The same equivalent circuit within the limit $\beta = \frac{R_{CM}}{R_N} \rightarrow 0$.

**Figure 6**: Relaxation time, $\tau$, as a function of the reciprocal temperature. Open squares are obtained from the fit (Equation (3)) and open circles from the Equation (1). The solid line is obtained by the product of the fitting functions describing $R_N(T)$ and C(T).

**Figure 7**: Permittivity of PCMO as a function of temperature.

**Figure 8** : Schematic drawing of: (a) Charge Ordering; (b) Orbital Ordering; (c) Dipole Ordering. The circles point the atomic positions and the crosses the charge ones.

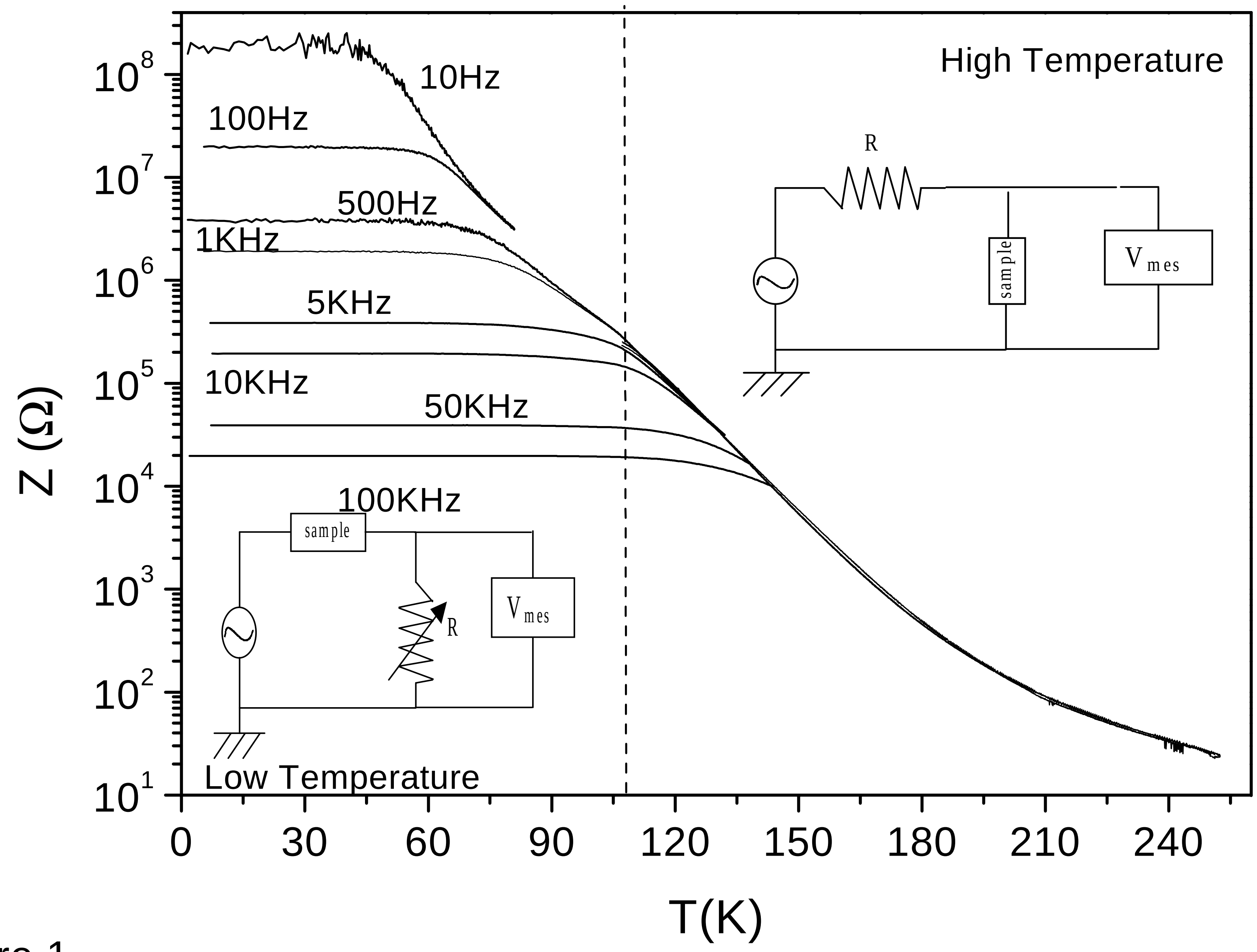

Figure 1
S.Mercone et al.

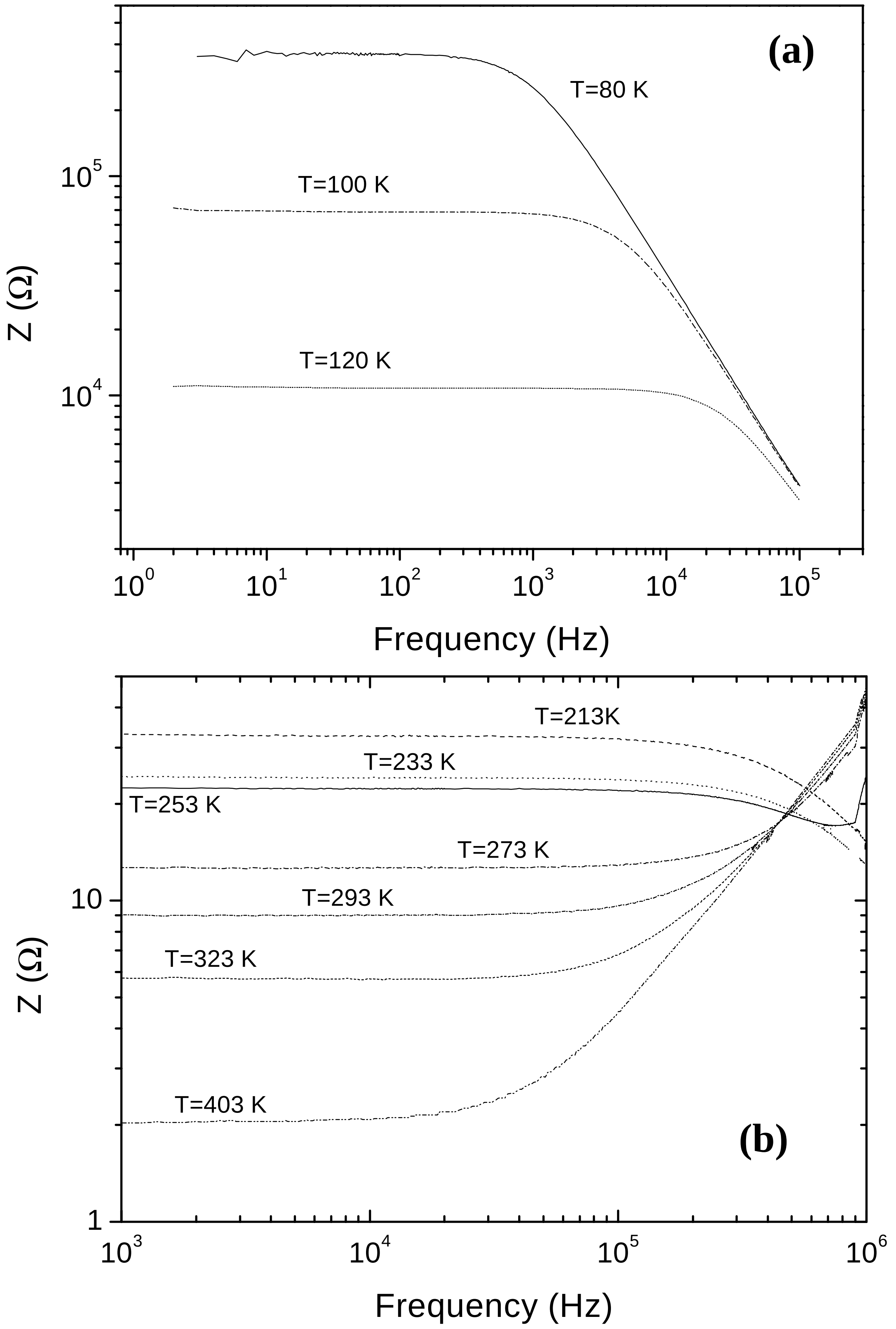

Figure 2
S.Mercone et al.

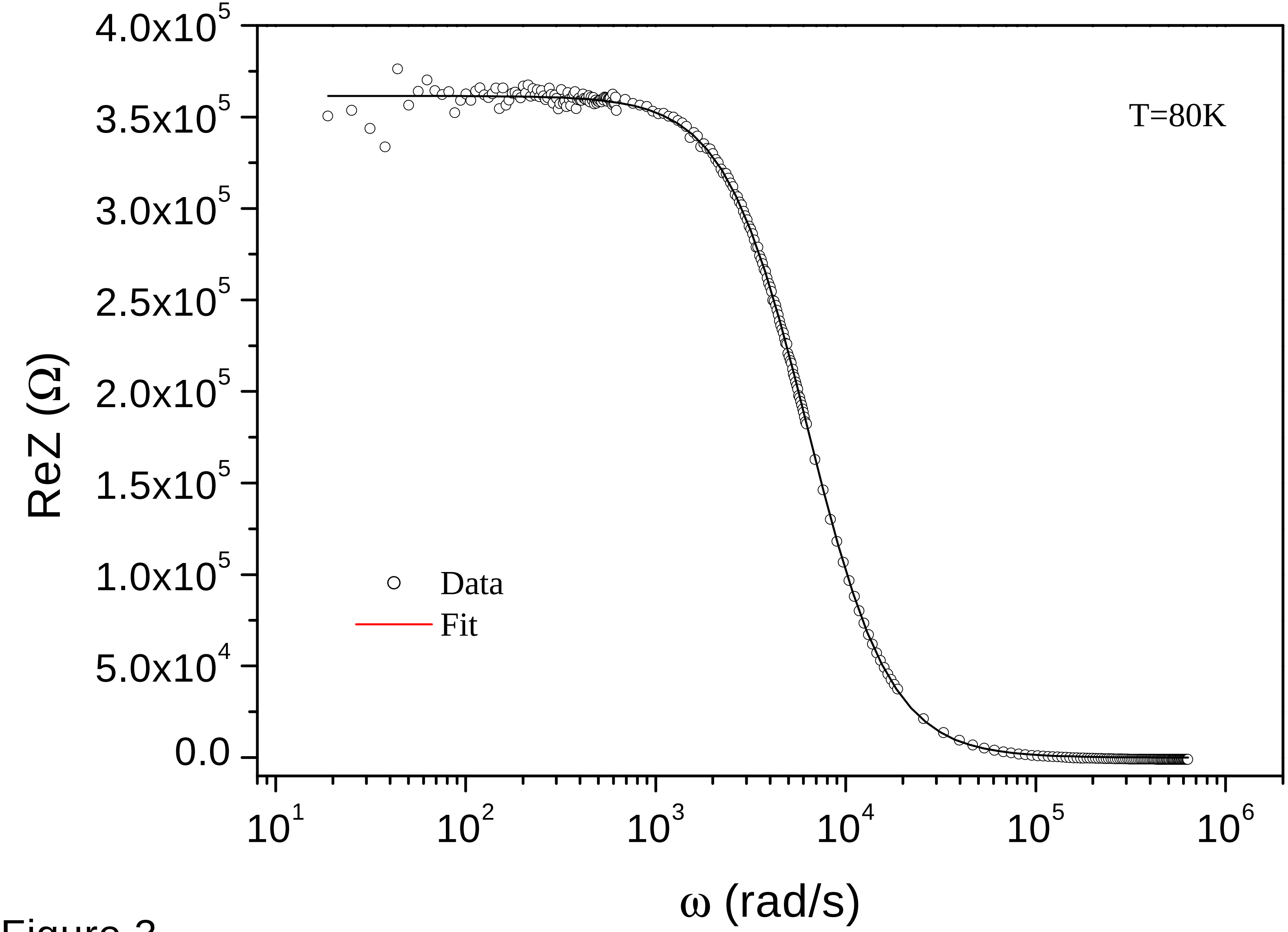

Figure 3
S.Mercone et al.

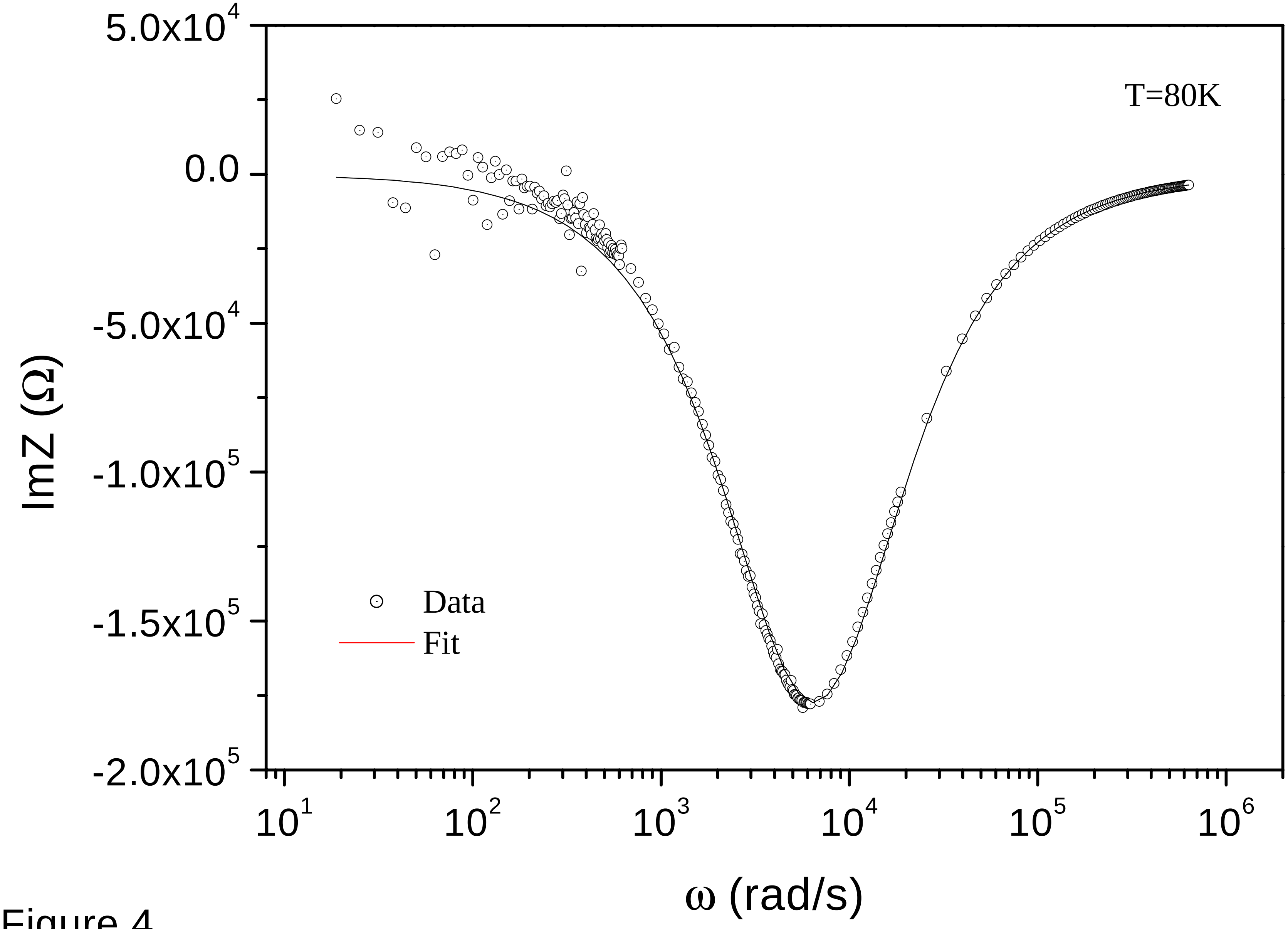

Figure 4
S.Mercone et al.

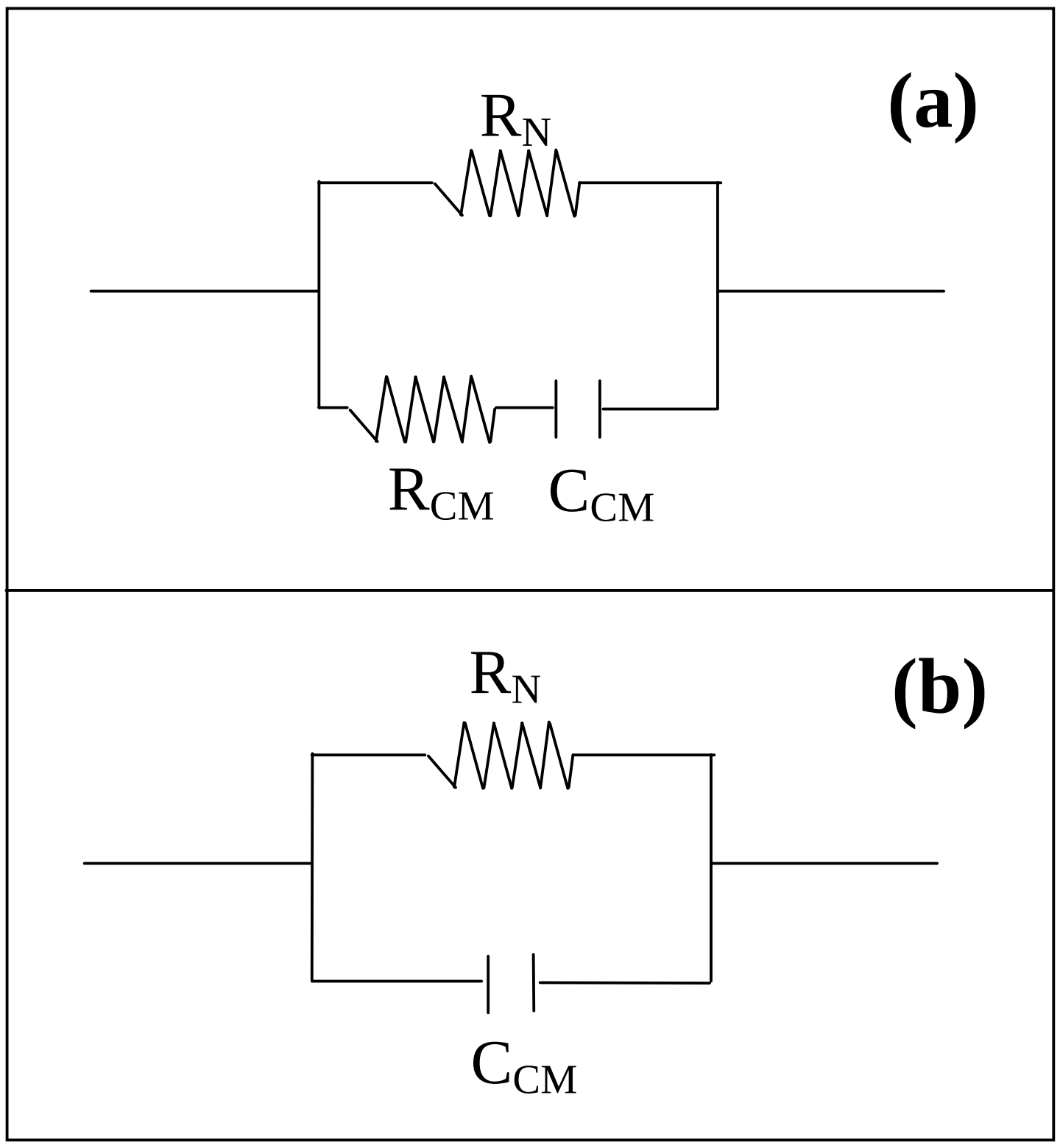

Figure 5
S.Mercone et al.

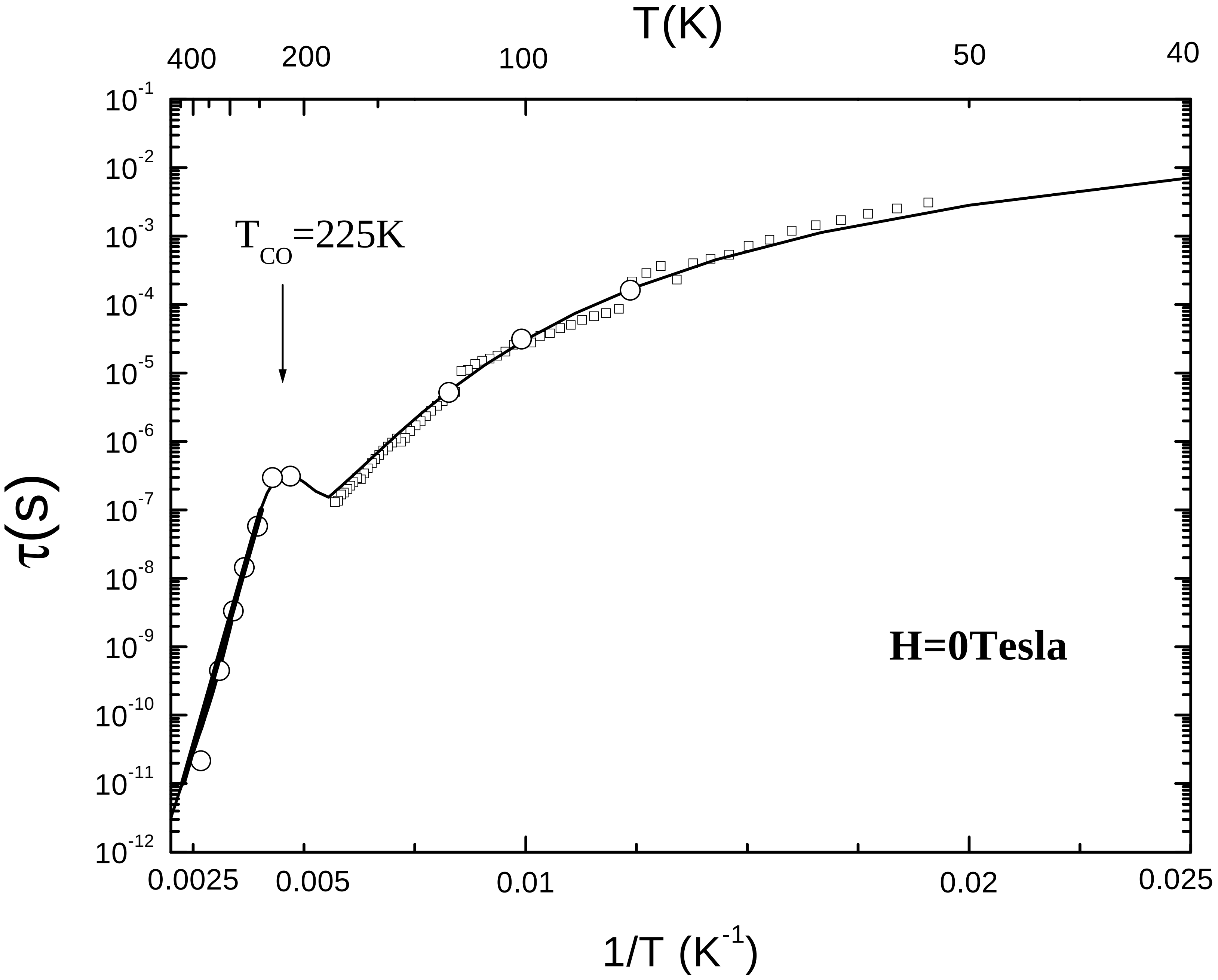

Figure 6
S.Mercone et al.

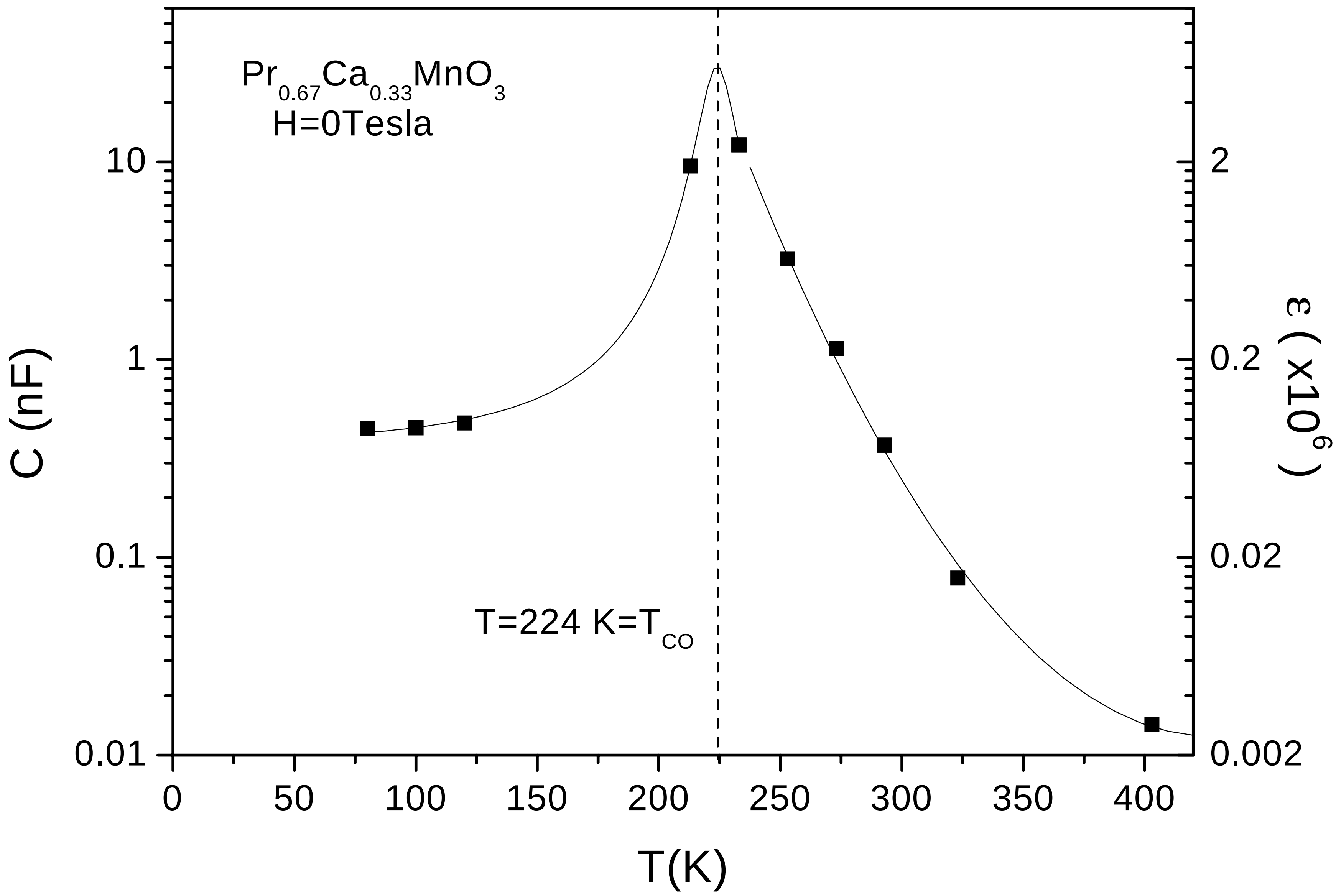

Figure 7
S.Mercone et al.